# A Novel Heart Disease Classification Algorithm based on Fourier Transform and Persistent Homology

Yin Ni
School of Mathematics and Statistics
Beijing Institute of Technology
Beijing, China
yinni@bit.edu.cn

Fupeng Sun
School of Mathematics and Statistics
Beijing Institute of Technology
Beijing, China
3120201455@bit.edu.cn

Yihao Luo
School of Mathematics and Statistics
Beijing Institute of Technology
Beijing, China
knowthingless@bit.edu.cn

Zhengrui Xiang
School of Mathematics and Statistics
Beijing Institute of Technology
Beijing, China
zx1321@imperial.ac.uk

Huafei Sun*
School of Mathematics and Statistics
Yangtze Delta Region Academy
Beijing Institute of Technology
Beijing & Jiaxing, China
huafeisun@bit.edu.cn

*Abstract*—Classification and prediction of heart disease is a significant problem to realize medical treatment and life protection. In this paper, persistent homology is involved to analyze electrocardiograms and a novel heart disease classification method is proposed. Each electrocardiogram becomes a point cloud by sliding windows and fast Fourier transform embedding. The obtained point cloud reveals periodicity and stability characteristics of electrocardiograms. By persistent homology, three topological features including normalized persistent entropy, maximum life of time and maximum life of Betty number are extracted. These topological features show the structural differences between different types of electrocardiograms and display encouraging potentiality in classification of heart disease.

## I. Introduction

According to the statistics of the World Health Organization, heart disease is one of the deadliest diseases in the world, threatening the health and safety of the world [1]. Diagnosing, predicting and preventing heart disease is therefore important for protecting lives and achieving personalized medicine in a growing population.

Electrocardiograms (ECG or EKG) record the electrical activity that causes the heart to contract and is used by medical professionals to diagnose heart disease [2]. The PQRST complex, core of electrocardiogram, is composed of waveforms that mark specific activity of the heart and can characterize the state of the heart [3]. In healthy ECG, the PQRST complex shows regular periodic variation.

For the diagnosis and prediction of heart disease, features extracted from single-lead and standard 12-lead ECG records have been analyzed in the literature. From the perspective of signal processing, I. Christov and C. Ye extract ECG signal shape, amplitude and other morphological characteristics [4, 5]; S. Banerjee and R. He uses wavelet transform to obtain characteristics in the frequency domain [6, 7]. From the perspective of dynamic system analysis, M. Richter regards the heart as a dynamic system, and applies delay embedding technology to phase space reconstruction of ECG signals for the first time [8]. In addition, the nonlinear dynamic system research methods such as Detrended Fluctuation Analysis (DFA) [9], Recurrence Quantification Analysis [10, 11] and Poincare Plot [12] have also been used to study ECG. From the perspective of statistics, T. Stamkopoulos obtains statistical features through higher-order statistics [13]. From the perspective of machine learning, researchers use SVM [14], convolutional neural network [7, 15], deep network [16], Boltzmann machine [17] and self-coding [18] to extract features. These perspectives and methods are often combined with each other in application, forming a deep yet incomplete research system.

Beyond the methods above, topological data analysis is a novel method to study ECG. As a widely used method for topology data analysis, persistent homology is an effective tool for data analysis because it provides concise, quantifiable, comparable, and robust summaries of the shape of data. Briefly, persistent homology reflects the changes of "holes" as simplicial complexes are constructed on point clouds. Persistent diagram and barcodes are the common outputs of persistent homology.

The periodicity of signals often leads to the constitutive property of point clouds, hence persistent homology has been successfully applied on the time series analysis of nonlinear dynamical systems to distinguish periodic signals from chaotic signals [19, 20, 21]. Time series analysis by persistent homology can be divided into four steps:

- Obtain point cloud from the original signal;
- Transform the point cloud into a simplicial complex with a high-dimensional skeleton structure;

- Reveal the geometric and topological characteristics of the whole point cloud by studying the persistent homology group of simplicial complex filtration and other topological invariants;
- Perform time series analysis based on topological features.

In lots of biological studies, the pathophysiology of physiological system is mostly manifested as the change from periodic to aperiodic, hence persistent homology shows its advantages in wheeze detection [22], pulse pressure wave analysis [23], mutational profiles in breast cancer [24], RNA structure analysis [25], spinal cord injury analysis [26] and other aspects of biological research.

Since pathological changes in heart are characterized by regular disruption of the PQRST complex in ECG, persistent homology used to be applied in ECG studies to classify different types of heart disease by analyzing ECG variations. The diversity of each step in the time series analysis using persistent homology results in the changes of research methods. At present, methods generating point cloud through ECG mainly include delayed embedding [27, 28], sliding windows embedding [29], ordinal partition graph [21] and sublevel set [30, 31]. The skeleton structures established on point clouds mainly include Vietoris-Rips (VR) complex [27, 29], Cech complex [28], ordinal partition graph [21] and sublevel set persistent [30, 31]. Common topological invariants extracted from persistent homology diagrams include 0-dimensional Betty number [31], 1-dimensional Betty number [27], time of life [28], the mean, variance and other statistics of barcode [29], normalized information entropy [21], topological triangle indices [30] etc. These topological features are then combined with machine learning by researchers to classify ECG.

In this paper, we construct the point cloud using sliding windows-fast Fourier transformation (SWFFT) embedding, which retains almost all ECG information due to the orthogonality of trigonometric functions. This is a novel attempt to transform an ECG into a point cloud and is first proposed by us. After transforming one ECG into point cloud by SWFFT embedding, VR complex is constructed.

Since VR complex shows the complete information of graph, while the information of other high-dimensional skeletons is incidental, we focus on obtaining 1-dimensional persistent homology diagram and barcode. Then three topological features, normalized persistent entropy, maximum life of time (max life) and maximum life of Betty number (max Betty life), were extracted. These topological features have been successfully applied to the classification of premature ventricular contraction, ventricular flutter, left bundle branch block and normal ECG.

The paper is organized as follows. Section II introduces background. In section III, Fourier persistent homology classification algorithm is proposed to classify different types of ECG and we will focus on how to construct point clouds using fast Fourier transformation and extract topological features from persistent diagrams. Section IV gives the results of classification and shows the accuracy. In section V, we draw the conclusion and discuss the future research direction.

## II. BACKGROUND

### A. Background of Three Types of Heart Disease

We mainly distinguish normal heart beat from premature ventricular contractions (P.V.C.), ventricular flutter and left bundle branch block.

P.V.C. arise from ectopic exciter in the ventricle and occur early in the heartbeat cycle, just before the next P wave. The ECG of P.V.C. shows very high and deep QRS complex wave. After P.V.C., there is a long compensatory interval, during which the heart is at rest. Generally, six P.V.C. per minute should be pathological.

Ventricular flutter is a fast rhythm whose ECG is a smooth sine wave. During ventricular flutter, the ventricles contract at a rate of five times per second, at the rate of which the coronary arteries cannot receive blood. Hence the heart itself has no blood supply. True ventricular flutter almost inevitably turns into ventricular fibrillation, requiring CPR and defibrillation.

Left bundle branch block is one kind of conduction delay. When one has left bundle branch block, his left ventricular excitation is slightly later than right ventricle. In the electrocardiogram, widened QRS complex wave (more than 0.12 seconds) and two R waves can always be observed. During these two R waves, the first part of wide QRS complex wave represents right ventricular depolarizing.

The normal ECG and three diseases ECG are shown in Fig.1.

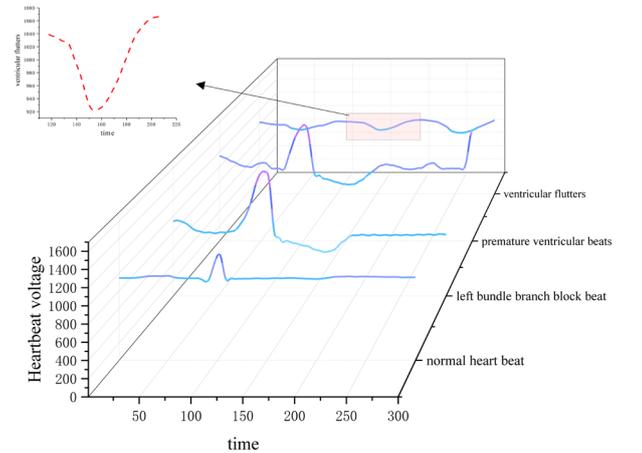

Fig. 1. *Four types of ECG*

### B. Fourier Transform and Fast Fourier Transform

Fourier transform is a method of signal transformation from time domain to frequency domain, allowing any signal to be resolved into waves with different frequencies. This transformation actually reconstructs signal in frequency domain, revealing certain features hidden in the time domain. However, traditional Fourier transform can only process continuous signal and what inputs into computer is discrete. As a result, we choose fast Fourier transform (FFT), which is a fully developed method for machine to compute spectrum of signal, whose computational complexity is $O(N \log_2 N)$.

The FFT separates the Fourier transform into even and odd indexed sub-sequences. When we input an n-points signal, the FFT will output a list of complex numbers:

$$\{L(i) \mid L(i) = a_i + b_i j, i = 1,2,3,...,n\} \quad (1)$$

The *i*-th point has several attributes:

- frequency: $F_i = (i-1)F_s / N$
- amplitude: $A_i = |a_i + b_i j|$ (2)
- phase position: $D_i = \text{atan}_2(b_i, a_i)$

where *Fs* is the sampling frequency and *atan2* is a function which returns the azimuth angle of origin to point $(x, y)$ ranging in $(-\pi, \pi)$.

The corresponding basis is

$$\{A_i \cos(2\pi F_i t + D_i) \mid i = 1,2,3,...,n\} \quad (3)$$

Notice that, if two signals share the same size *N* and *Fs*, the output lists will have the same frequency lists. Hence their corresponding basis are same, which provides stable criterion for producing point cloud.

*C. Simplicial Homology*

As a kind of algebraic operators from topological spaces to groups, simplicial homology groups are abelian groups which can classify different topological spaces. For a topological space *X*, the idea of simplicial homology is to consider *X* as a space gluing by convex polyhedral of different dimensions and study the "hole" in *X*. The simplicial homology groups of *X* can be obtained as follows:

Step1: Construct the $\Delta^n$-complexes structure

$$\{\sigma_\alpha : \Delta^{n(\alpha)} \to X, n(\alpha) \in \mathbb{Z}_{\geq 0}\}_{\alpha \in J} \quad (4)$$

satisfying three properties [34], where $\Delta^n$ is a standard *n*-complexes.

Step2: Define the simplicial chain group of *X*:

$$\Delta_n(X) = \underset{\alpha, n(\alpha)=n}{\mathbb{Z}} \sigma_\alpha = \left\{ \sum_{\alpha, n(\alpha)=n} \lambda_\alpha \sigma_\alpha \mid \lambda_\alpha \in \mathbb{Z} \right\} \quad (5)$$

where $\lambda_\alpha$ are almost all zero.

Step3: Define the chain map (boundary map)

$$\partial_n : \Delta_n(X) \to \Delta_{n-1}(X) \quad (6)$$

via $\alpha$ such that $n(\alpha) = n$ and

$$\partial_n(\sigma_\alpha) = \sum_{i=0}^{n} (-1)^i \sigma_\alpha \Big|_{[v_0, \cdots, \hat{v}_i, \cdots, v_n]} \quad (7)$$

It is straightforward calculation to check $\partial_n \partial_{n+1} = 0$.

Step4: Calculate the simplicial homology group of *X*

$$H_n^\Delta(X) = \ker \partial_n / \text{Im}\, \partial_{n+1} \quad (8)$$

Elements in $\ker \partial_n$ and $\text{Im}\, \partial_{n+1}$ are called *n*-cycles and *n*-boundaries respectively.

The dimension of $H_n^\Delta(X)$ is called *n*-th Betty number, denoted by $\beta_n$. Betty number can describe the number of "holes" of a topological space. For instance, the 0-th Betty number counts the connected components, the 1-th Betty number represents the number of holes and the 2-th Betty number computes numbers of voids.

*D. Persistent Homology*

Persistent homology is one of the methods in TDA and searches structural differences of data. The idea of persistent homology is to watch how the simplicial homology changes over the course of a given filtration.

Given dimension *n*, if there is an inclusion map *i* of one topological space *X* to another *Y*, then it induces an inclusion map on the *n*-dimensional simplicial chain groups

$$i : \Delta_n(X) \to \Delta_n(Y) \quad (9)$$

Furthermore, this extends to a homomorphism on simplicial homology group

$$i_* : H_n^\Delta(X) \to H_n^\Delta(Y) \quad (10)$$

where $i_*$ sends $[c] \in H_n^\Delta(X)$ to the class in $H_n^\Delta(Y)$.

Hence, the persistent homology of a point cloud *P* is constructed as follows.

Step 1: Transform point cloud *P* into a topological space

A common method is to construct VR complex. For given $r \geq 0$ and metric *d* in *P*, the VR complex $VR(P, r)$ is the topological space containing different dimensional simplex whose maximum distance among vertices is less or equal than $2r$.

Step 2: Construct a filtration of topological spaces.

A filtration $X_1 \subseteq X_2 \subseteq ... \subseteq X_m$ induces a sequence of homomorphisms on the simplicial homology groups

$$H_n^\Delta(X_1) \to H_n^\Delta(X_2) \to ... \to H_n^\Delta(X_m) \quad (11)$$

A class $[c] \in H_n^\Delta(X_i)$ is said to be born at *i* if it is not in $i(H_n^\Delta(X_{i-1}))$. The same class dies at *j* if $[c] \neq 0 \in H_n^\Delta(X_{j-1})$ but $[c] = 0 \in H_n^\Delta(X_j)$.

Take VR complex as an example, a filtration of VR complex can be obtained with the growing of *r*. By changes of *r*, Fig.2 shows that new rings emerge and old rings die.

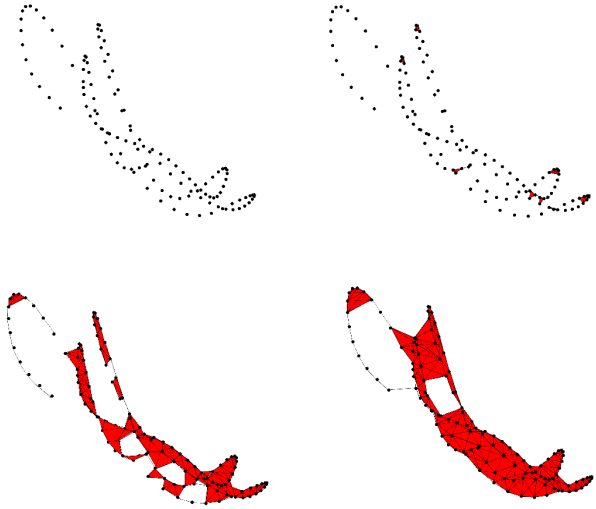

Fig. 2. A filtration of Vietoris-Rips complex

Step 3: Obtain the persistent information

We mark a point in $\mathbb{R}^2$ at $(i, j)$ if one class is born at $i$ and dies at $j$. Hence, we can acquire a persistent diagram by its collection of off-diagonal points

$$D = \{(b_1, d_1), \ldots, (b_k, d_k)\} \quad (12)$$

The lifetime or barcode of a point $x = (b, d)$ in $D$ is given by $\text{pers}(x) = |b - d|$.

Drawing barcode is a convenient way to visualize the lifetime of each class. Fig.3 is an example of barcode of a point cloud generated from normal ECG, showing the life time of detected features.

In addition, there exists another presentation for the information of the persistent homology, called the persistent diagram, where we directly draw pairs in $D$ onto $\mathbb{R}^2$. The examples of persistent diagrams are presented in Fig.6.

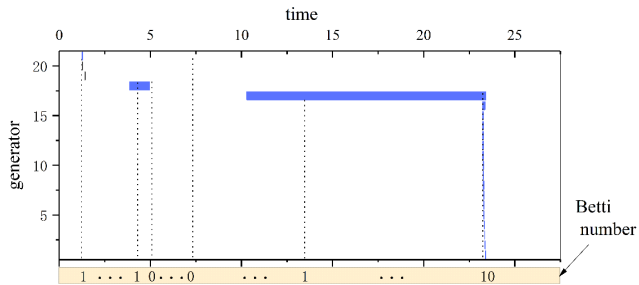

Fig. 3. Barcode of a point cloud generated from normal ECG

## III. FOURIER PERSISTENT HOMOLOGY CLASSIFICATION ALGORITHM

### A. Pipeline of the Algorithm

As shown in Fig.4, we adopt an innovative pipeline of topological data analysis to classify the mixed ECG data.

The algorithm has four steps:

Step 1: Preprocessing of ECG signal.

We adopt the strategy of Butterworth filter algorithm to cut off noisy portions with spectral power over 50 Hz, develop a local search algorithm to find the periodic R-peak and successfully segment continuous ECG into single heartbeat.

Step 2: Sliding Window Fast Fourier Transform (SWFFT).

SWFFT is a mapping that maps the signal into point cloud. Let $S = \{s_i\}_{i=0}^{n-1}$ be the signal of length $n$ and $F$ be the function of FFT. Given window length $d$ and sliding speed $\tau$, the signal $S$ is divided into $m = \left\lfloor \dfrac{n-d}{\tau} \right\rfloor$ child-signals:

$$P_{\tau,d}(S) = \{p_k\}_{k=0}^{m-1} \quad (13)$$

where

$$p_k = [s_{k\tau}, s_{k\tau+1}, s_{k\tau+2}, \ldots, s_{k\tau+(d-1)}] \quad (14)$$

FFT is then carried out on $P_{\tau,d}(S)$. Let $F(p_k)$ be the output of FFT, which is a sequence of complex numbers

$$F(p_k) = [a_k^1 + b_k^1 j, \ldots, a_k^d + b_k^d j] \quad (15)$$

Hence, we can calculate the expansion coefficient of child-signal $p_k$ under the same basis, marking as $Pc(p_k)$. From equation (2) and equation (3), we have

$$Pc(p_k) = [A_k^1 \cos(D_k^1), \ldots, A_k^d \cos(D_k^d), \\ A_k^1 \sin(D_k^1), \ldots, A_k^d \sin(D_k^d)] \quad (16)$$

Since each child signal in $p_k \in P_{\tau,d}(S)$ shares the same size $d$ and sampling frequency *Fs*, their output sequences share the same basis. By this way, we can embed $P_{\tau,d}(S)$ into the point cloud $Pc(P_{\tau,d}(S))$ by transforming each child signal into a point in $\mathbb{R}^d$.

Step 3: Persistent homology

Having the point cloud $Pc(P_{\tau,d}(S))$, we use VR complex to establish the corresponding topological structure.

With persistent homology, we can obtain persistent diagram and barcode. Based on these conditions, three topological features, normalized persistent entropy, max life and max Betty life, are extracted to reveal the changes of periodicity of signals.

Step 4: Classification

In 3-dimensional topological feature space, each signal corresponds a topological feature vector. We use a simple SVM

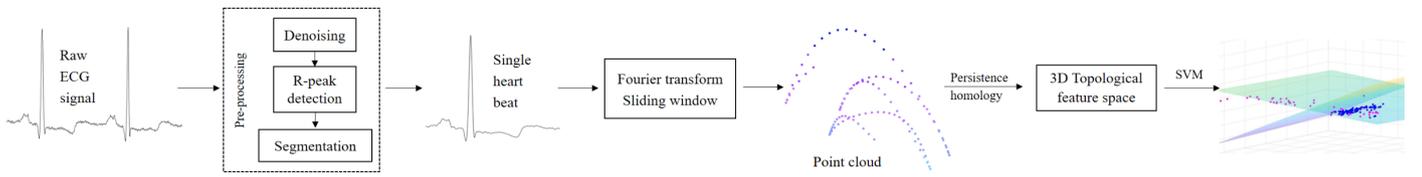

Fig. 4. Pipeline of topological data analysis to classify the mixed ECG

classifier to get segmentation plane set $\pi$, which would be the criterion to differentiate signals.

*B. Case study*

For each kind of heart beat, the ECG, point cloud and persistent diagram are visualized in Fig.6. Through SWFFT, different ECG are turned into point clouds with different structures. In order to describe the persistence information of cycle, we pay attention to generators with lifetime longer than 0.05. These disparities are enlarged by means of persistent homology.

The 1-dimensional Betty number is a kind of topological invariant, marking the scales of 1-dimensional cycles in different topological space. Hence, in the process of constructing graph, we select the max Betty life to describe the most stable situation. Furthermore, by examining the persistence diagram, we find the distance from points to diagonal different, which represents the life time of each generator. Thus, we use max life to express the most lasting cycle. This method is currently mainstream [32, 33].

In addition to focusing on details of persistent diagram, we need to describe the overall property. Entropy is then introduced to revel chaoticity and disorder. However, traditional definition of entropy will cause deviation as a result of few significant lifetime. Thus, we choose the persistent entropy

$$E'(D) = \frac{E(D)}{\log_2(\mathcal{L}(D))} \quad (17)$$

$$E(D) = -\sum_{x \in D} \frac{\text{pers}(x)}{\mathcal{L}(D)} \log_2\left(\frac{\text{pers}(x)}{\mathcal{L}(D)}\right) \quad (18)$$

where, the $\mathcal{L}(D)$ is the sum of lifetimes of points.

From Fig.6, we notice that, the ventricular flutter has the minimum max Betty life, which means the complex generated by its point cloud is more unstable. This information coincides with the shape of its point cloud. The P.V.C. owns the longest max life, meaning that there are 1-dimensional generators lasting long time. This phenomenon may result from the unique arrangement of P.V.C. point cloud, which is a representation of its tall and deep QRS integrated wave. From Fig.5, healthy ECG has a more concentrated distribution of persistence entropy whose center is 2. It inspires us that a violation of ECG regularity can result in a fair-sized increase or decrease in entropy.

IV. EXPERIMENTS

*A. Data*

The sample data set comes from the MIT-BIH Arrhythmia Database, which is one of the first generally standard arrhythmia test databases. The database consists of 48 half-hour ambulatory ECG recordings from 47 subjects. After the preprocessing, to prove our model effective, 100 healthy heart beats, 50 left bundle branch block beats, 50 premature ventricular beats and 50 ventricular flutters are chosen randomly, making up the sample data set.

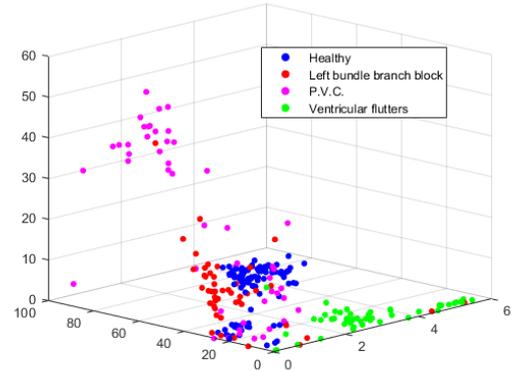

Fig. 5. Distribution of 4 kinds ECG in feature space

*B. Simulation*

We develop Fourier transform persistent homology algorithm code module through python (entire code and database can be found on https://github.com/bitni2609/FFT-persistent-homology-ECG). After carrying out the module on sample data set, the corresponding 3-dimensional topological feature space is constructed. Fig.5 shows the distribution in feature space. The SVM is then introduced to get the set of segmentation plane.

Two typical SVM methods, one of which has 'Linear' kernel and another has 'Gaussian' kernel, are implemented to output plane set $\pi_1$ and hyperplane set $\pi_2$ respectively. The 'Linear' kernel SVM has regularization parameter C, which controls the loss and makes model more precise. Apart from C, the 'Gaussian' SVM also owns kernel coefficient $\gamma$, which makes model closer to support vector. In order to decrease the overfitting impact, we set C and $\gamma$ to be 0.4 and 0.5 respectively.

TABLE I. BINARY CLASSIFICATION OF HEALTHY ECG

| Heart disease | kernel | |
|---|---|---|
| | *Linear* | *Gaussian* |
| Ventricular flutters | 0.99 | 1.00 |
| Left bundle branch block | 0.82 | 0.99 |
| P.V.C. | 0.85 | 0.97 |

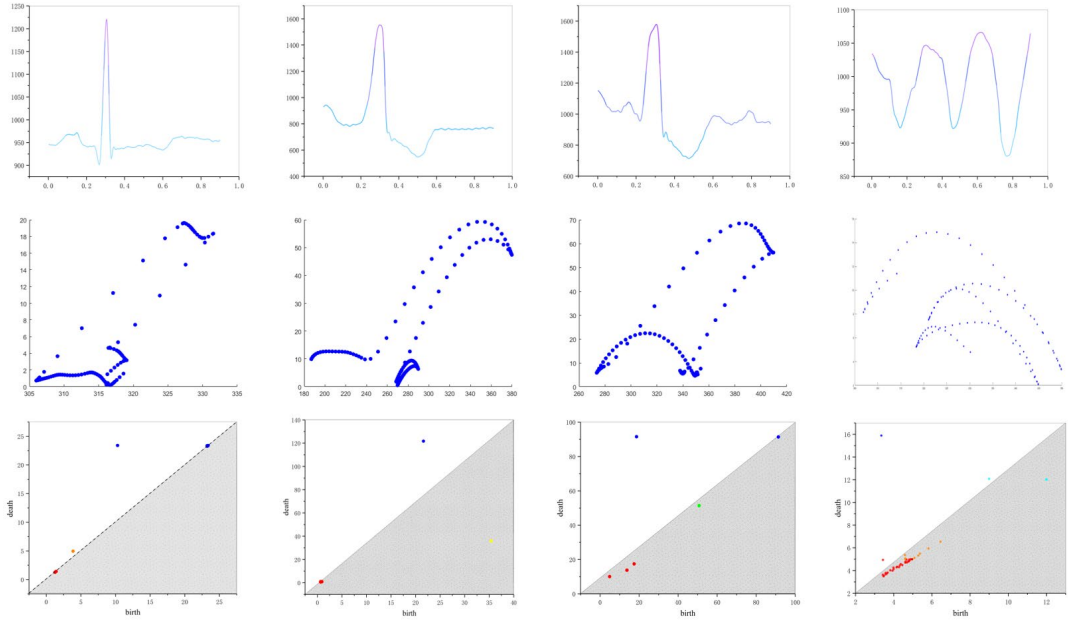

Fig. 6. The figures from the top row to the bottom row present four types of ECG, point cloud of ECG and persistent diagrams of point cloud respectively. The first column to the last column presents the case of normal ECG, P.V.C., left bundle branch block and ventricular flutter respectively.

Accuracies of distinguishing healthy ECG with the other three cardiac diseases respectively are shown in Table I.

For any of the three diseases, our model is capable to differentiate between healthy and the cardiac heartbeats with accuracy of over 82%. If we use SVM with 'Gaussian' kernel, the average accuracy can reach over 98%. This result proves that our model can effectively distinguish healthy ECG with cardiac diseases.

When we inspect Fig.5, it shows that there are certain intersections in the feature space, especially for healthy and left bundle branch block. But the 'Gaussian'-kernel SVM can 100% distinguish healthy signal, which shows the existence of overfitting. In other words, there are small zones in $\pi_2$ wrapping marginal healthy ECG points. This phenomenon is not beneficial for more expansive sample test. Besides that, certain anomalous segmentation zones are not visible and oddity-from sample. Hence, 'Gaussian'-kernel SVM is not suitable for further classification.

We then test classifying four kinds of ECG with 'Linear' kernel SVM. The accuracy is shown in Table II.

TABLE II. CLASSIFICATION ACCURACY OF FOUR ECG

| Heart disease | kernel |
|---|---|
| Healthy | 0.97 |
| Ventricular flutters | 0.96 |
| Left bundle branch block | 0.50 |
| P.V.C. | 0.50 |

We draw how SVM with "Linear" kernel segments in Fig.7. From Table II and Fig.7, the segmentation of healthy and unhealthy is highly effective, but the accuracy of distinguishing P.V.C and Left bundle branch block is barely satisfactory. Fig.6 revels the residual. These two diseases have similar ECG, which causes the inefficiency of SWFFT to extract topological features. However, the result still proves our model effective in classifying ECG signal.

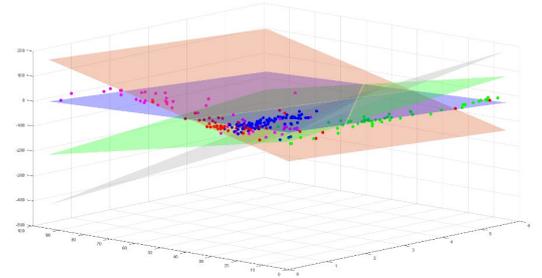

Fig. 7. Final classification based on three topological features

## V. CONCLUSION AND FUTURE WORK

This paper presents a new method of topological analysis for ECG classification. By sliding windows-fast Fourier transform (SWFFT) embedding, ECG is transformed into point cloud, and normalized persistent entropy, max life, and max Betty life were extracted. In a 3-dimensional topological feature space, we use SVM to classify four types of ECG. In actual application scenarios, the plane set may change as the data set changes. But our experimental results show the effectiveness of our algorithm, especially for distinguishing ventricular flutters in small sample set.

ACKNOWLEDGMENT

This research was funded by National Key Research and Development Plan of China, No.2020YFC2006201. Special thanks to Mr. Xingming Gao for his luminous suggestions.